\documentclass{aastex}
\usepackage{spr-astr-addons}
\usepackage{latexsym}
\usepackage{amssymb,amsbsy}
\usepackage{amsmath}
\usepackage{amsfonts}
\usepackage{multirow}
\usepackage{rotating}
\usepackage{graphicx}
\usepackage{epsfig}
\usepackage{color}
\usepackage{natbib}

\RequirePackage{color}
\newcommand\viz {$viz.,\;$}

\newcommand\abt {$\sim$}
\newcommand{\msun}{M_{\odot}}
\newcommand{\mbh}{M_{\bullet}}

\newcommand{\msigma}{M_{\bullet}-\s}

\def\barr{\begin{array}}
\def\earr{\end{array}}
\def\berr{\begin{eqnarray}}
\def\err{\end{eqnarray}}
\def\berrno{\begin{eqnarray*}}
\def\errno{\end{eqnarray*}}
\def\be{\begin{equation}}
\def\ee{\end{equation}}
\def\fr{\frac}
\def\la{\langle}
\def\ra{\rangle}

\def\barr{\begin{array}}
\def\earr{\end{array}}
\def\berr{\begin{eqnarray}}
\def\err{\end{eqnarray}}
\def\berrno{\begin{eqnarray*}}
\def\errno{\end{eqnarray*}}
\def\be{\begin{equation}}
\def\ee{\end{equation}}

\def\fr{\frac}
\def\la{\langle}
\def\ra{\rangle}

\renewcommand{\tt}{transformation }

\newcommand{\pol}[1]{\stackrel{\rm LCP}{\mathrm{RCP}}}

\def\apjs{{\it Astrophys.~J.~Suppl.}}

\def\apjl{{\it Astrophys.~J.~Lett.}}

\def\aap{{\it Astron.~Astrophys.}}

\renewcommand{\a}{\alpha}
\renewcommand{\b}{\beta}

\newcommand{\e}{\epsilon}

\renewcommand{\o}{\omega}
\renewcommand{\O}{\Omega}

\newcommand{\s}{\sigma}

\shorttitle{BH correlations for globular clusters}
\shortauthors{Safonova and Shastri}
\begin{document}

\title{Extrapolating SMBH correlations down the mass
scale: the case for IMBHs in globular clusters}
 
\author{Margarita Safonova\altaffilmark{1} and Prajval Shastri\altaffilmark{1}}
\affil{Indian Institute of Astrophysics,  Sarjapur Road, Bangalore 560 034}

\altaffiltext{1}{rita@iiap.res.in}
\altaffiltext{2}{pshastri@iiap.res.in}

\begin{abstract}

Empirical evidence for both stellar mass black holes ($\mbh <10^{2}\,\msun$) and supermassive 
black holes (SMBHs, $\mbh>10^{5}\,\msun$) is well established. Moreover, every galaxy with a 
bulge appears to host a SMBH, whose mass is correlated with the bulge mass, and even 
more strongly with the central stellar velocity dispersion $\sigma_c$, the $\msigma$ relation.
On the other hand, evidence for "intermediate-mass'' black holes (IMBHs, with masses 
in the range 100--$10^{5}\,\msun$) is relatively sparse, with only a few mass measurements 
reported in globular clusters (GCs), dwarf galaxies and low-mass AGNs. We explore the question 
of whether globular clusters extend the $\msigma$ relationship for galaxies to lower black 
hole masses and find that available data for globular clusters are consistent with the 
extrapolation of this relationship. We use this extrapolated $\msigma$ relationship to 
predict the putative black hole masses of those globular clusters where existence of
central IMBH was proposed. We discuss how globular clusters can be used as a constraint 
on theories making specific predictions for the low-mass end of the $\msigma$ relation.

\end{abstract}

%

\section{Introduction} 
 
The empirical evidence for the ubiquity of both stellar mass black holes (black hole mass $\mbh$ 
of ~\abt~$1-15\,\msun$) and supermassive black holes (SMBHs, $\mbh$~$>10^{5}~\msun$) 
is well established. While it is estimated that there are about $10^7-10^9$ 
stellar-mass black holes in every galaxy \citep[e.g.,][]{ShapiroTeukolsky83,BrownBethe94}, 
every galactic bulge appears to host a SMBH. Moreover, the kinematically determined mass 
of these SMBHs is correlated with the mass of the bulge, and even more strongly with the 
central stellar velocity dispersion $\sigma$, the $\msigma$ relation \citep{FerrareseMerritt00,gebhardtetal2000c,KormendyGebhardt01,Tremaine02}. Active galactic 
nuclei (AGN), which have long been believed to be driven by accretion around SMBHs and 
whose SMBH masses have been estimated from reverberation-mapping, are also consistent 
with the $\msigma$ relationship \cite[e.g.,][and references therein]{Wandel02,Bentz09}.
The corollary is that the nuclear activity is a phase (or more) in the life of (at least) 
every galaxy with a bulge. 

The $\msigma$ relation has been extended down to black hole masses of \abt~$10^5\,\msun$ 
\citep{Barth05,GreeneHo06} by searching for central BHs within very low-luminosity AGN. Their 
dynamical studies are an unambiguous verdict on the presence of central BHs in dwarf 
ellipticals and very late-type spirals (e.g., NGC4395 by \citet{GreeneHo07} and references 
therein). Even lower mass black holes have been inferred from non-dynamical methods for other 
low-luminosity AGN \cite[e.g.,][]{Dong07}. In spite of considerable 
efforts, however, evidence for black holes of still lower mass, \viz the intermediate-mass black 
holes (IMBHs, of masses $10^2-10^4\msun$), is relatively sparse. Attempts to discover IMBHs 
in globular clusters via their X-ray emission have been on for a long time, 
\cite[e.g.,][]{BahcallOstriker75}, and although the recently discovered ultra-luminous 
X-ray sources (ULXs) have been attributed to IMBHs \cite[e.g.,][]{ColbertMushotsky99}, 
this suggestion has also been contested \cite[e.g.,][]{Berghea08}. 

Extrapolating the $\msigma$ correlation down the mass scale predicts that IMBHs 
can be found in stellar systems that have velocity dispersions of $<30$~km/sec, 
clearly pointing to the globular clusters. Observational evidence for such black holes 
are only a handful, however ({\it cf.} references in Table~\ref{tab:sampleclusters}). 
Furthermore, theoretical results on IMBHs in globular clusters remain ambiguous, at best.
Some theories predict the necessity of most (if not all) Galactic 
globular clusters to host central black holes \citep[e.g.,][]{MillerHamilton02}, 
while some argue for the impossibility of globular clusters to form 
and/or retain black holes in their cores \citep[e.g.,][]{Favata04,KawakatuUmemura05}. 
The importance of investigating the possibility of globular clusters hosting IMBHs  
cannot be overestimated. While a central IMBH would clearly impact the evolution 
of the globular cluster itself, more generally, IMBHs are crucial to link the 
formation processeses of stellar-mass BHs and SMBHs, and could have served as 
seeds for the growth of SMBHs. Extending the local BH mass function to the extreme 
low end can help in understanding whether there is a minimum galaxy mass or 
velocity dispersion below which BHs are unable to form or grow \citep[e.g.,][]{Bromley04}. 
Globular clusters and galactic bulges are both `hot' stellar systems, and, since all large bulge 
systems seem to have a central black hole, to the extent that a massive, bound 
globular cluster can be viewed as a ``mini-bulge", it may be that every dense 
stellar system (small or large) hosts a central black hole \citep{Gebhardt02}. 
BHs at the low end of the mass ladder can be used as a constraint on theoretical 
models with different predictions on the $\mbh-\s$ behaviour. For example, the 
prediction that $\msigma$ relation shall substantially steepen below $\s=150$ km/sec 
\citep{Granato04} is not supported by the low-mass AGN sample \citep{Barth05}. 
The question of IMBHs also has bearing on debates in cosmology, since the cosmic 
mass-density of IMBHs could exceed that of SMBHs ($\O\approx 10^{-5.7}$), and 
may even account for all the baryonic dark matter in the universe, with 
$\O\approx 10^{-1.7}$ \citep{vanderMarel03}.

The few recent reports on the detection of central black holes 
in some globular clusters seem to suggest that globular clusters do 
follow the $\msigma$ relation for SMBHs \citep{vanderMarel02,McLaughlin06,
Noyola08,Lanzoni07,Gebhardt02} and that these BHs represent the `true' IMBHs in 
the mass range of $10^2-10^4\,\msun$. In this paper, we compile published discoveries 
of central black holes in globular clusters and investigate the consistency of 
the data with the $\msigma$ and $\mbh-$luminosity relations. Using these 
globular clusters as a constraint on the slope of the extended $\msigma$ 
relation, we estimate the black hole masses for a sample of globular 
clusters that are proposed to host IMBHs and discuss the implications.

\section{$\mbh-\s$ Correlation}
\label{sec:Msigma}

Although the $\msigma$ relationship for galaxies is very well established, the 
published estimates for the slope parameter $\b$ of the relationship span a
wide range (3.68 to 4.86), the reasons for this discrepancy discussed 
by \cite{Tremaine02} and \cite{FerrareseFord05}. As a rough estimate from
this correlation, for a typical globular cluster with cenral dispersion 
$\s_c$ of the order $10$ km/sec, the mass of the central black hole 
would be $\mbh \sim 10^3\,\msun$. This roughly coincides with the estimate 
from the Maggorian relation $\mbh$\abt~10$^{-3\pm 1}\,M$ \citep{Magorrian98}, 
and with the theoretical estimates of the formation of the central BH from the 
initial $\sim 50 \msun$ seed by stellar/low-mass objects accretion 
\citep{MillerHamilton02}, $\mbh$=10$^3\,\msun\times 10^{-(M_{\rm V}+10)/2.5}$.

\subsection{Constraints on the $\msigma$ relation by globular clusters candidates} 
\label{subsec:sigma_fits}

For investigation of the very low-end extrapolation of the $\msigma$
relation, we take the globular clusters where the presence of the central 
black hole was inferred from optical observations of either individual stars or by 
using integrated light techniques. The data on the clusters are given in 
Table~\ref{tab:sampleclusters}.

\begin{table*}
\begin{center}
\caption{Globular clusters with the reported central black holes}
\label{tab:sampleclusters}
\vskip 0.1in
\begin{tabular}{|c|c|c|c|c|c|}
\hline
\rule[-3mm]{0mm}{8mm}
Cluster &Other    & $\mbh$ & $\s$     & $M_{\rm V}$ & Ref.\\
Name    &  Name   &  ($10^3\,\msun$)  & (km/sec) &    & \\ 
\hline
\rule[-3mm]{0mm}{8mm}NGC~104          & 47Tuc   &$1.0^{+0.5}_{-0.5}$  &$11.6\pm 0.8$&-9.42 &1\\  
\rule[-3mm]{0mm}{8mm} NGC~5139$^{*}$   & $\o$ Cen&$40.0^{+7.5}_{-10.0}$&$23.0\pm 2.0$&-10.29&2\\ 
NGC6388$^{\dag}$                      &         &$5.7^{+5.7}_{-2.85}$ &$18.9\pm 3.6$&-9.42 &3\\
\rule[-3mm]{0mm}{8mm}NGC~7078$^{\ddag}$& M15     &$2.5^{+0.7}_{-0.8}$  &$14.1\pm 3.2$&-9.17 &4,5,6 \\
\rule[-3mm]{0mm}{8mm} G1 (M31)        & Mayal II&$18.0^{+5.0}_{-5.0}$ &$25.1\pm 1.7$&-10.94&7\\
\hline
\end{tabular}
\end{center}
\noindent
\vskip 5mm
{\it Key to columns:} (1)-(2) cluster identification number and name,
(3) BH mass $\mbh$ with error bars (where reported), 
(4) central velocity dispersion, (5) V-band absolute magnitude, (6) references are the BH 
measurement papers: (1) \citet{McLaughlin06}; (2) \citet{Noyola08}; 
(3) \citet{Lanzoni07}; (4) \citet{vanderMarel02}; (5) \citet{Gerssen02}; (6) \citet{Gerssen03}; 
(7) \citet{Gebhardt05}.\\
$^{*}$ this value has recently been reduced to the upper limit of $1.2\times 10^4\,\msun$ 
\citep{AndersonvanderMarel09}, see discussion in Sec.~\ref{subsec:regions}.\\
$^{\dag}$ a factor of two uncertainty has been assigned to this BH mass.\\
$^{\ddag}$  the reported masses for this cluster range from $1$ to 
$9\times 10^3\,\msun$; the listed mass is the most probable value, see latest \citep{Kiselev08}.
\end{table*}  

\begin{figure*}[ht!]
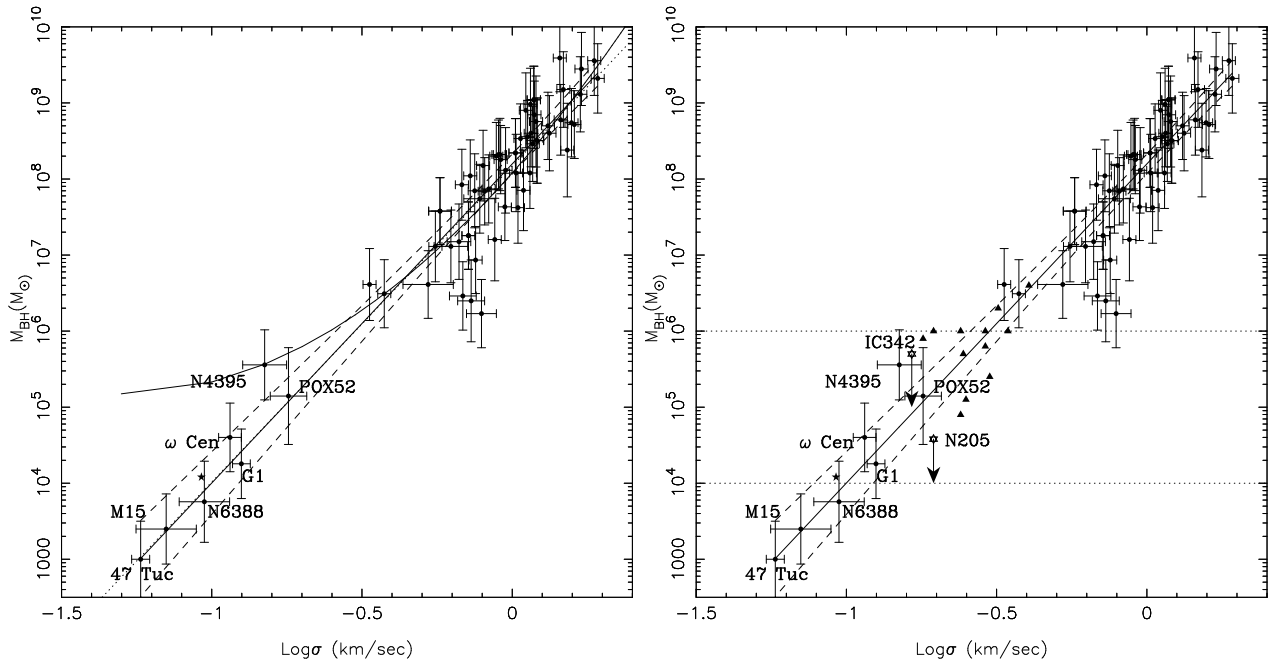

\begin{center}
\epsfig{figure=SafonovaShastri_fig1.ps,width=0.5\textwidth,angle=-90}
\epsfig{figure=SafonovaShastri_fig2.ps,width=0.5\textwidth,angle=-90}
\caption{\small Mass of the central black hole $\mbh$ (in solar masses) 
vs. velocity dispersion $\s\,(=\s_e/200)$ km/sec for galaxies and globular 
clusters. Dotted line is the linear
regression fit for only galaxies and solid line is the fit including the globular
clusters. Dashed line is the 2-sigma error on the regression. Solid curved line is 
the quadratic fit for GRG sample (see in the text). Single star is the most 
recent estimate of the $\omega$~Cen black hole mass \citep{AndersonvanderMarel09}. 
{\it Right}: Clearly seen the division of the plot into three distinct regions,
upper part is occupied by SMBHs, middle by MBHs with inclusion of N205 and IC342 
upper limits, bottom by IMBHs. Filled triangles are the sample of low-mass 
AGNs of \cite{GreeneHo06}. See discussion in Sec.~\ref{subsec:regions}.} 
\label{fig:msigma}
\end{center}
\end{figure*}

In Figure~\ref{fig:msigma} we plot the black hole masses and central velocity dispersions for 
globular clusters (Table~\ref{tab:sampleclusters}), together with the sample of galaxies 
with secure BH mass estimates available to date. The galaxy data consist mainly of a 
sample of 49 galaxies \citep{Gultekin09} with directly dynamically-measured black hole 
masses (called GRG sample hereafter). The galaxies included in their sample are only 
elliptical or with classical bulges or psedobulges, and since a globular cluster can be 
viewed as a `mini-bulge', we favour this sample. To the GRG sample we have added six 
galaxies from \cite{Hu08}; the reason for their absence in GRG sample is unclear, 
since their mass measurements are also dynamical, and the galaxies are either elliptical 
or having pseudobluges. We have also added the latest estimates of the black hole mass 
in NGC~4395 \citep{Peterson05} and Pox 52, the lowest-mass AGN to date \citep{Barth04}. 
Though mass estimates for these two black holes are not from gas or stellar dynamics 
(the reason why AGN, for example, were excluded from GRG sample), these measurements 
were made by multiple methods and are considered to be quite reliable. The data for 
these additional galaxies are given in Table~\ref{tab:addgalaxies} in Appendix~B.

We assume that there is an underlying relation of the form
\be
y=\a +\b x\,,
\label{eq:basic}
\ee
where $y=\log(\mbh/\msun)$, $x=\log(\s_e/\s_0)$; $\s_0$ is some reference 
value usually chosen to be 200 km/sec (it was noticed by \cite{Tremaine02} 
that this choice reduces the uncertainties in the intercept $\a$), and $\s_e$ 
is the effective velocity dispersion, defined in \citet{Gultekin09} as the spatially
averaged, rms, line-of-sight stellar velocity within the effective radius $r_e$. It was 
noticed by \citet{MerritFerrarese01} that  $\s_e$ might be expected to reflect the 
depth of the stellar potential well more accurately than the central velocity dispersion, 
$\s_c$. However, when this was not available, we have used $\s_c$ as, on average, there 
is remarkably little difference between $\s_e$ and $\s_c$ \citep{MerritFerrarese01,Gultekin09}. 
Many of the galaxies in the sample have assymetric quoted errors, but 
for simplicity we assume here that the measurement errors are symmetric in both 
parameters with root-mean-square (rms) values $\e_{xi}$ and $\e_{yi}$ for galaxy 
(or cluster) $i$. In our sample we used the published bounds to the 1$\s$ range in 
BH mass, and upper and lower limits to the dispersion\footnote{\footnotesize 
Errors in $x$ and $y$ are assumed to be $(\log{\s_{\rm upper}}-
\log{\s_{\rm low}})/2$ and $(\log{M_{\bullet,\rm high}}-
\log{M_{\bullet,\rm low}})/2$, respectively. The measurement errors 
in mass are added together with the intrinsic scatter, $\e_0=0.4$, 
in quadrature, using the best estimate of $\e_0$ from GRG 
sample.}. The goal is to estimate the 
best-fit values of $\a$ and $\b$ and their associated uncertainties for the 
total sample of galaxies and five globular clusters (using the symmetric 
least-squares method).

Figure~\ref{fig:msigma} shows the correlation between $\mbh$ and the velocity
dispersion of the host. The best-fit linear relation for the sample with only 
galaxies (designated ``Subsample") is
\be 
\log\left(\fr{\mbh}{\msun}\right)= ( 8.17\pm 0.06) + (4.16\pm 0.3)
\log\left(\fr{\s_e}{\s_0}\right)\,,
\label{eq:fit1}
\ee
and the slope of the relation changes only slightly with inclusion of the 
five globular clusters (``Full sample")
\be 
\log\left(\fr{\mbh}{\msun}\right)= (8.19 \pm 0.06) + (4.20\pm 0.2)
\log\left(\fr{\s_e}{\s_0}\right)\,.
\label{eq:fit2}
\ee

These data are summarised and compared with the best results from \citet{Gultekin09} 
in Table~\ref{tab:msigma_stats}.

\begin{table*}
\begin{center}
\caption{}
\label{tab:msigma_stats}
\vskip 0.1in
\begin{tabular}{|l|c|c|c|c|c|}
\hline
Sample & $\a$  & $\b$  & $\chi^2$   &  $\chi^2/$dof  \\
\hline
Subsample & $8.17\pm 0.06$  & $4.16\pm 0.3$  & 50.9 & 0.85 \\
Full sample &$8.19\pm 0.06$ & $4.20\pm 0.2$ & 50.9 & 0.78 \\ 
GRG (TO2ind)$^{\dag}$ & $8.19\pm 0.06$ & $4.06\pm 0.37$ & ... &... \\
GRG best$^{\ddag}$    & $8.12\pm 0.08$  & $4.24\pm 0.41$ & ... & ...\\  
\hline
\end{tabular}
\end{center}
\vskip 5mm
\noindent
$^{\dag}$ Method of \cite{Tremaine02} applied to a GRG sample without the
upper limits \citep{Gultekin09}. The intrinsic scatter is added in quadrature 
to the measurements errors in the expression for $\chi^2$. \\
$^{\ddag}$ The best fit in \citet{Gultekin09}. 
\end{table*}

The fact that globular clusters fit the $\msigma$  relation is remarkable, and
rather puzzling, considering that the formation mechanisms for the central black
holes are believed to be different: the SMBHs in galaxies are hypothesized
to have formed through gas collapse and subsequent gas accretion, 
while the central black holes in globular clusters are believed to have formed 
through either collapse of a stellar cluster \citep[e.g.,][]{PortegiesZwartMcMillan02} 
or through stellar (or low-mass remnants) accretion onto an
existing relatively massive black hole \citep{MillerHamilton02}. 
However, both theories agree that 
there may have been an initial seed, an IMBH of order $\sim 10^3 \msun$
for the case of galaxies, and a $\sim 50\,\msun$ black hole for the 
case of globular clusters. The fact that all proposed BHs in globular 
clusters fit the $\msigma$ correlation suggests that there must be some 
previously unrecognized connection between the formation and evolution 
of globular clusters, galaxies and their central black holes.

\subsection{SMBHs to IMBHs through massive black holes}
\label{subsec:regions}

Apart from the remarkably consistent extrapolation of the $\msigma$ relation down
to the very low end, the right panel in Figure~\ref{fig:msigma}
displays three distinct regions. The top right corner is populated by a well-known and
well-established sample of a SMBHs, which has a relatively sharply defined
low edge. It was proposed long ago that there might be a physically 
determined lower limit [$>10^6\msun$] to the mass of a SMBH \citep{Haehnelt98}, 
and recently \citet{WehnerHarris06} argued that formation of a black hole is 
strongly favoured above this limit. It was the need to determine the low
end of the SMBH mass function that started the active IMBH search. Elucidating 
the demographics of low-mass BHs can provide the critical 
input to test the theoretical models of quasar formation \citep{Haehnelt98}.
In particular, the low-mass cut-off in the BH mass function today 
provides a constraint on the mass function of seed BHs. Models of 
different seed masses predict different BH occupation fractions at the low 
end of the mass ladder, ranging from some galaxies hosting tiny 
($<10^4$~$\mbh$) BHs to virtually no central black holes below 60 km/sec (see discussion in 
Sect.~4.2).
 
Dynamical studies present an unambiguous verdict on the existence of central black 
holes in dwarf stellar systems, and these BHs can be properly named massive 
black holes --- MBHs --- as opposed to their higher-mass counterparts, 
with a range in masses of $\sim 10^4-10^6\msun$ and velocity dispersions 
of $\sim 25-65$ km/sec. This region is flanked from the top by a few massive/supermassive 
galaxies such as, for example, M32, Milky Way and Circinus with the lowest mass in 
SMBH region of $1.7\times 10^6\msun$. It is populated by  well-established dwarfs,
such as NGC~4395 and POX52 (see Table~\ref{tab:addgalaxies}). The upper limits 
for black holes in the galaxies IC342 and NGC~205 ($1.5\times 10^4 \msun$ from 
\citet{Boeker99} and $2.2\times 10^4\msun$ from \citet{Valluri05}, respectively) 
lie in this region as well. However, current dynamical techniques do not have 
the spatial resolution needed to resolve the gravitational sphere 
of influence of a low-mass ($\sim 10^5\msun$) BH outside of the Local Group, while even 
local studies can only place upper limits (e.g. NGC~205), or reject the existence of 
a BH altogether (M33). Therefore, the search is forced to 
use AGN activity as a signature for the BH. The recently discovered SDSS DR4 sample of 
dwarf Seyfert 1 nuclei galaxies, referred to herein as the sample of \cite{GreeneHo06}, 
may represent an upper envelope of the population of MBHs --- black holes in low-mass 
galaxies. This sample has a median mass of $\la\mbh\ra=1.3\times 10^6\msun$ 
\citep{GreeneHo06} with the lowest BH mass of $8\times 10^4\msun$, and the 
host galaxies do belong to the low-mass (low-luminosity) population. 
The broad-line AGN SDSS DR4 sample with mass range of $5\times 10^4-10^6\msun$ 
\citep{Dong07} brings the present census of these AGN MBHs up to 400. 
However, since the majority of this MBH population are AGNs, being quite 
compatible with the low-end extension of active SMBHs in their properties, 
the question still remains as to whether normal dwarfs also have central BHs. 
Furthermore, globular clusters G1 and $\o$~Cen, that lie at the lower end of the MBH 
region, are now believed to be not genuine globular clusters, but  
the cores of the accreted galaxies stripped by tidal forces 
\citep[][and refs therein]{MackeyvandenBergh05}. As such, the central 
black holes of G1 and $\o$~Cen are still {\it galactic} black holes, 
and their masses of $\sim \mbox{few} \times 10^4\,\msun$ thus represent 
the lowest limit for the massive {\it galactic} black holes. Incidentally, 
the ratio of $\mbh$ to the total mass of $\o$~Cen is 
much higher than the ``canonical" value of $\sim 0.3\%$, but if $\o$ Cen is the core 
of an accreted dwarf, most of its mass would have been stripped during the tidal 
evolution and, once corrected for that, a mass of $4\times 10^7\,\msun$ obtained 
by modeling \citep{Noyola08}, puts the observed black hole near the $0.3\%$ value, 
which suggests that when this galaxy was accreted, it already 
possessed a central black hole. It should be noted that recently \citet{AndersonvanderMarel09} 
reported new, reduced, values for both BH mass and velocity dispersion leave the 
$\omega$ Cen black hole on the same $\msigma$ relation, and in the same MBH region.

The other, ``genuine", globular clusters, such as M15, 47 Tuc and NGC~6388, represent 
a third region of the `genuine' IMBH domain, narrowing the mass range of IMBHs to 
$\sim 10^2-10^4\msun$.

Thus, the earlier existing gap in the BH `mass ladder' seems to be rapidly 
filling up; with BHs smoothly populating the whole range from $1 \msun$ up to nearly 
$10^{10}\msun$, with a clear distinction of mass ranges 
between different stellar systems; stellar-mass black holes are loners 
or in binary systems, IMBHs live in small stellar systems like GCs, 
and massive black holes (MBH+SMBH) in the cores of galaxies. This
brings back the question of dynamical detection of BHs in low-mass non-active
stellar systems, with globular clusters being the best potential candidates. 
Most of the Galactic globular clusters are close by, and while optical observations 
are still difficult due to the extreme crowding in the centres, it may be 
possible to discover BHs from radio and X-ray observations 
and, in the future, from microlensing \citep{SafonovaStalin06} 
and gravitational waves's observations.

\subsection{Black hole mass estimates in a sample of globular clusters}
\label{subsec:mass_estimates}

It was found that the $\msigma$ relation is a powerful tool as it allows the 
prediction of BH masses (which are difficult to measure directly) from 
readily available galaxy parameters \citep{Lauer07}; for example, from a 
single, low-resolution observation of a galaxy's velocity dispersion, its central BH mass 
can be estimated with an accuracy of $\sim 30\%$ or better \citep{FerrareseMerritt00}. 
Notably, BH masses measured by reverberation-mapping 
and BH virial mass estimates for some local AGNs are broadly consistent 
\citep{gebhardtetal2000c,GreeneHo06}. Based on the remarkable consistency of 
the BH masses discovered in globular clusters described in the previous section, 
we applied the $\msigma$ relation to estimate the 
masses of possible central black holes in some Galactic globular clusters. 
Although the precise nature of this correlation is not yet understood, there may 
be no limit to the range of velocity dispersions over which the $\msigma$ relation can
apply \citep{BegelmanNath05}. We have applied the results of the Sec.~\ref{subsec:sigma_fits}
(Eq.~\ref{eq:fit2}) to estimate the masses of black holes for a sample of globular 
clusters and presented the results in Table~\ref{tab:GCmasses}. The list of candidate 
globular clusters grows continuously as different methods of detections are being 
suggested. Our sample of Galactic GCs that may host a central black hole 
(Table~\ref{tab:GCmasses}) was compiled on the basis of these different proposals.  

Since high-centrally concentrated globular clusters and especially 
core-collapsed clusters have long been proposed as harbouring central 
black holes, we have included in our sample all Galactic core-collapsed 
(or post-core-collapsed) clusters \citep{Trager95}. However,
this view has been challenged recently by \cite{Baumgardt05} who has argued 
that, on the contrary, no core-collapsed clusters can harbour central black 
holes, as they would quickly puff up the core by enhancing the rate of close 
encounters, tending to prevent the core collapse and leaving an imprint of a 
shallow cusp; in other words, the clusters with flat large cores and King 
profiles outside would rather host a central black hole. This view was further 
reinforced by \cite{Miocchi07}, who in addition suggested looking for 
the extreme horizontal-branch (EHB) stars as a possible fingerprint of 
the presence of an IMBH. However, \citet{Hurley07} issued a cautionary
note on using large core radius as an indicator of an IMBH presence, since
other factors, such as the presence of a stellar BH-BH binary in the
core, can flatten the measured luminosity profile and enlarge the core radius,
and that it is still too early to abandon the IMBH indicators used earlier, 
such as the steepening of a M/L ratio in GC cores \citep{Gebhardt05}. In our 
candidate sample (Table~\ref{tab:GCmasses}) we have included 
candidates from \cite{Baumgardt05} and \cite{Miocchi07}.  

It should be noted that in Table~\ref{tab:sampleclusters} of reported 
BH cases both variants are included, G1 and $\o$ Cen having large cores and 
M15 a collapsed core. NGC~6388 is the cluster for which both \cite{Miocchi07} 
and \cite{Baumgardt05} agree on the existence of the central black hole. 
 
Recently \citet{DrukierBailyn03} have suggested that high-velocity 
stars in GCs cores can be used to reveal the central BHs and have 
drawn a list of most probable clusters based on existing observations. 
Based on radio emission as a possible test of the existence of IMBHs 
in GCs, \citet{Maccarone04} suggested that the radio, rather than the 
X-ray, observations can be a more successful test for a central 
IMBH and have calculated expected radio flux densities for a few possible candidates. 
One more technique to search for the massive BH-BH binary system in the core 
of GCs is the use of exotic ejected binary systems, for which the cluster NGC~6752 
is a very good candidate, having a binary pulsar system at 3.3 
half-mass radii from the core and a high central mass-to-light ratio 
\citep{Colpi03}. In Table~\ref{tab:GCmasses}, the last but one 
column lists the reason for the selection of a cluster. Black hole masses 
from these different predictions are listed in the last column of the Table.

\section{$\mbh$ -- Luminosity correlation}

The black hole mass in galaxies correlates linearly with the absolute  
luminosity of the bulge of the host galaxy \citep{KormendyRichstone95,Magorrian98}.
We use here the extinction-corrected,  bulge (or total for ellipticals), V-band 
luminosities calculated from absolute magnitudes, using
\be
\log{(L_V/L_{\sun,V})}=0.4 (4.83-M_V)\,,
\ee
where $M_V$ for the GRG sample is taken from \citet{Gultekin09}, for additional galaxies
from \cite{Lauer07} (see Table~\ref{tab:addgalaxies}), and for globular clusters from
the Harris catalog \citep{Harris96}. By the same fitting method as in Sec.~\ref{subsec:sigma_fits},
with the exception of using measurement errors only on mass 
\cite[routine `fit' from][]{Press92}, the linear regression for our sample gives 
\be
\log\left(\fr{\mbh}{\msun}\right)=(8.89\pm 0.08) + (1.03\pm 0.04) \log\left(\fr{L_V}{10^{11} L_{\sun,V}}\right)\,.
\label{eq:M_luminosity}
\ee
Comparison with other fits is given in the Table~\ref{tab:M_luminosity}. 

\begin{table*}
\begin{center}
\caption{}
\label{tab:M_luminosity}
\vskip 0.1in
\begin{tabular}{|l|c|c|c|c|c|}
\hline
Sample & $\a$  & $\b$  & $\chi^2$   &  $\chi^2/$dof  \\
\hline
Full sample &$8.89\pm 0.08$ & $1.03\pm 0.04$ & 68.7 & 1.6 \\ 
Subsample & $9.05\pm 0.10$  & $1.33\pm 0.11$  & 62.5 & 1.52 \\
GRG best$^{\ddag}$    & $8.95\pm 0.11$  & $1.11\pm 0.18$ & ... & ...\\  
Tremaine sample$^{\dag}$ & $8.41\pm 0.11$ & $1.40\pm 0.17$ & ... &... \\
\hline
\end{tabular}
\end{center}
\vskip 5mm
\noindent
$^{\dag}$ Symmetric least-squares fit applied to the \cite{Tremaine02} sample
by \cite{Lauer07}. \\
$^{\ddag}$ The best fit in \citet{Gultekin09}. 
\end{table*}

\begin{figure}[h!]
\centerline{
\epsfig{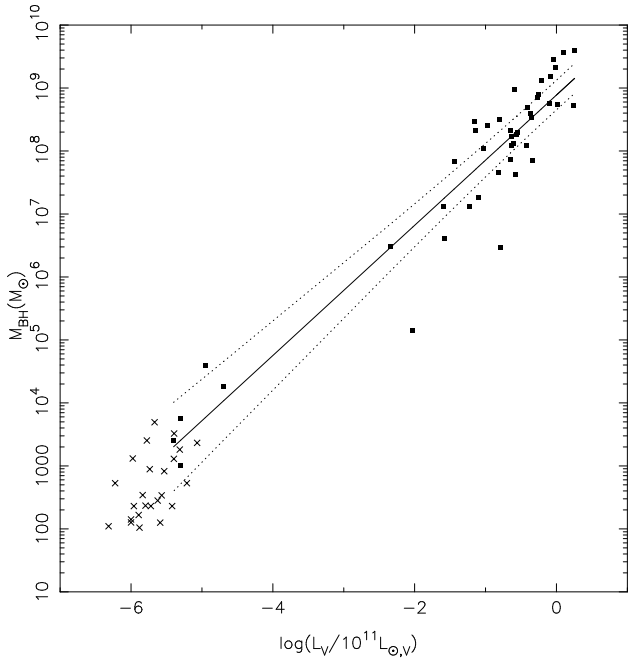}}
\caption{\small Mass of the central black hole $\mbh$ (in solar masses) 
vs. V-band luminosity of the bulge (or cluster) for galaxies and globular 
clusters. The original full sample is represented by the filled squares. The
clusters from Table~6 are represented by crosses. Solid line is the linear 
regression fit for our full sample. Dashed lines are 3-sigma errors on the 
regression fit. Clusters with masses inferred from Eq.~\ref{eq:fit2} (crosses) 
fall on and around the best-fit line.} 
\label{fig:m_luminosity}
\end{figure}

Since five reported globular clusters fit this correlation quite well (within the scatter),
we may expect that the sample of candidate globular clusters will follow the same. We 
have plotted the calculated black hole masses for globular clusters from
the Table~\ref{tab:GCmasses}, denoted by crosses, against their V-band luminosities in
Fig.~\ref{fig:m_luminosity}. It should be noted from Table~\ref{tab:GCmasses} 
that black hole masses depend strongly on velocity dispersion, with 
$\s_c \lesssim 7$ km/sec giving a mass too low to be considered seriously. 
It is most likely that only the black holes of masses $\gtrsim 100 \,\msun$ 
would remain at the centres, with the lighter ones ejected from the cluster due 
to gravitational interactions \citep[e.g.,][]{MillerHamilton02}. 
Thus, for the Fig.~\ref{fig:m_luminosity} we retained only the black holes 
with masses $> 100\,\msun$. The solid line gives the best fit for our full sample
and the dashed lines give the 3-$\s$ confidence limits on the fit. 

It seems intriguing that the $\mbh$--luminosity relation is preserved even for the
putative GC black holes. The larger scatter than for $\msigma$ relation is also
observed, just like for the more massive counterparts. Since the $\mbh$ -- luminosity relation
is considered to be the reflection of $\mbh$ -- bulge mass relation, and GCs follow
different fundamental-plane relations than galaxies with classical bulges  
\citep[e.g.,][and the numerous references cited therein]{WehnerHarris06}, it seems
strange that they shall continue on the same relation. At least for the
low-mass local AGN sample, \citet{Greene08} suggested that this relation will look 
different, with a bias towards more massive host galaxies for a given BH mass. This relation
is also suspected to deviate in the regime of the most luminous elliptical galaxies, 
with black hole masses of more than $10^9\,\msun$ \citep{Lauer07}. It was proposed 
that, at that end, the bulge luminosity is a better predictor for a black hole mass 
than velocity dispersion, whereas at the low-mass end (globular clusters), 
the BH growth really depends on a (relatively) local stellar velocity dispersion, 
as would be expected in a model where the BH accretion is fed by the capture of 
individual stars \citep{MiraldaEscude05} and, moreover, the GC mass is susceptible 
to the tidal erosion. When we applied the weighted least-squares fit 
(neglecting measurement errors in both variables for simplicity, using 
routine `fit' from \citet{Press92}) to the sample of only globular clusters 
(five clusters from Table~\ref{tab:sampleclusters} plus candidate clusters from 
Table~\ref{tab:addgalaxies}), the slope indeed increased, but not dramatically, 
only to $1.355\pm 0.25$, approaching the slope for the sample of only galaxies. 
Similar analysis was done by Miocchi (2007) for the sample of
globular clusters, where black hole masses were estimated by
constraining the parametric model of a globular cluster with IMBH from
observed central surface density profile and concentration parameter.
In their sample the black hole mass appears to be poorly correlated with
the V-band luminosity of a cluster, however it shall be noted that
black hole masses obtained by that method have huge uncertainties and
the sample used to obtain the correlation is quite small. Their
$\msigma$ relation also has a very shallow slope of 1.2. When using 
the $\msigma$ relation for the globular cluster 
sample, though, it should be borne in mind that the reported velocity 
dispersion measurements for most of our candidate clusters may not be 
representative of the (higher) actual values in the inner (unresolved) 
regions of the clusters. Observations of globular clusters with higher 
spatial resolution may provide better constraint on this low-end mass-ladder extension.
The latest example of under-estimation of the central velocity dispersion is the report
by \cite{Ibata09}, in which the globular cluster M54 (NGC~6715) is found to 
host a $\sim 10^4\,\msun$ black hole; the previous central velocity dispersion 
from, for example, \cite{PryorMeylan93} was $\s=14.2$ km/sec, 
but \cite{Ibata09} report a value of  $\s=20.2$ km/sec.

\section{Discussion}

\subsection{Central BHs and evolution of globular clusters} 

The fact that globular clusters fit the extrapolation of the $\msigma$ and 
$\mbh-$luminosity correlations for massive black holes raises the possibility that
there may be some previously unrecognized connection between the formation and 
evolution of galaxies, globular clusters and their central black holes. 

Just as the correlation between the BH mass and bulge properties of the galaxies 
sheds light on their formation and evolution histories, the existence of BH in 
globular clusters may help in understanding the evolution of globular clusters, 
and how significantly, for instance, might the evolutionary effects influence the 
survival of a cluster with a central IMBH. 

\begin{figure}[h!]
\begin{center}
\includegraphics[width=90mm]{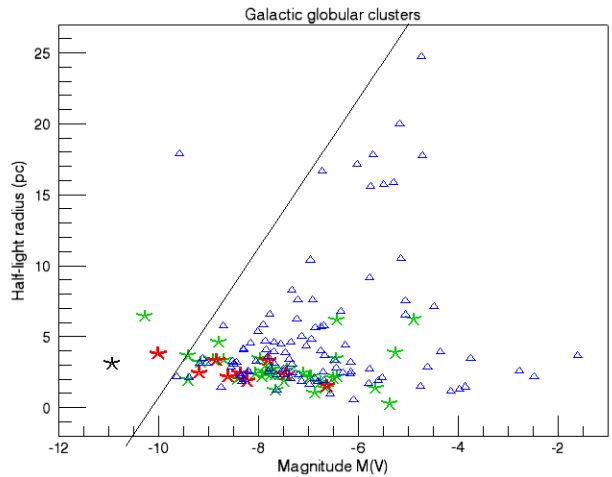}
\caption{$R_{h}$ vs $M_V$ for 146 Galactic globular clusters (data from 
\citet{MackeyvandenBergh05}). The
Baumgardt's set is marked by red asterisks; the CC set by green
asterisks; the remaining clusters by blue triangles. The M31
globular cluster G1 is marked by a black asterisk, the measurement
for its $R_{\rm h}$ is taken from \cite{Barmby02}.
Above the slanted line the red asterisk marks $\o$ Cen, the green
asterisk M54.} 
\label{fig:2}
\end{center}
\end{figure}

Fig.~\ref{fig:2} shows $R_{\rm h}$, the radius that contains half of 
the cluster stars in projection, plotted against $M_V$, the integrated luminosity, 
for 146 Galactic globular clusters, highlighting the separate sets of GCs by 
different symbols. The candidate set of \cite{Baumgardt05} is denoted by green 
asterisks, the CC and post-CC candidates by red asterisks and the remaining 
Galactic clusters are marked by blue triangles. Most theories agree that 
present-day IMBHs most probably reside in massive, dense and concentrated 
globular clusters, for example, the ones that
occupy the region near and under the straight line in Fig.~\ref{fig:2}, rather than
in loose and dissolving clusters, i.e., clusters from the so-called 
`graveyard' \citep{MackeyvandenBergh05} --- the rightmost region in Fig.~\ref{fig:2}. 
There is a
sharp edge to the main distribution of the clusters \citep[the Shapley line,][]{vandenBergh08b}, 
and only three Galactic clusters ($\o Cen$, M54 and NGC~2419), 
and an M31 cluster G1 (marked by a black asterisk), lie above the upper envelope. 
All of these globular clusters, including the rather 
diffuse NGC~2419 (the uppermost left blue triangle), are believed to be the stripped cores of 
now-defunct dwarf spheroidals accreted by our Galaxy. Though the recent work
by \cite{vandenBergh08a} indicates that this plot may be not the best way to 
distinguish between globular clusters and dwarf spheroidals, (for example, 
NGC~2419 may still be a genuine globular cluster, see refs in \citet{vandenBergh08a}), 
it confirms $\o Cen$, M54 and G1(M31) to be ex-dwarfs on the basis of their 
ellipticity. Several clusters, {\it viz.,} 47~Tuc, M15, NGC~6388 and NGC~6441, which is one of 
the four brightest Galactic globular clusters, lie very close to this line.  
The clustering of our candidates in a central area of the plot, in the region where
the clusters that are both `tight and bright' lie, may indicate 
that dense and high-luminosity clusters are better candidates for central black 
hole searches than diffuse and low-luminosity ones. 

While some have argued on theoretical grounds that a central IMBH in the core 
may increase the stability
of the cluster and thus its efficiency to withstand the tidal disruption for a longer
time during its passages through the Galaxy \citep[e.g.,][]{Hansen03}, the contrary 
view also exists, that a central IMBH speeds up the dissolution of a star cluster, especially 
if a cluster is surrounded by a tidal field \citep{Baumgardt04}. The observational 
consequences of the latter view are that there may exist a population of `rogue' 
IMBHs in the Galaxy (as MACHOS, for example), and that if such a cluster spirals 
into the Galactic centre, the result would be an IMBH with a group of stars still 
tidally bound to it (for example, it can explain the existence of a stellar complex
IRS~13E within 1~pc of the Galactic centre as the remains of a GC with an IMBH 
\citep{Hansen03}). However, isolated IMBHs as MACHOs would have been detected by 
microlensing surveys, which, incidentally, have ruled out the existence of MACHOs
in the mass range of $10^{-8}-100\,\msun$, and recently, at the 95\% confidence
level, MACHOs with masses $M > 43 \,\msun$ were excluded at the standard local halo density
\citep{Yoo04}, thus closing the MACHO mass window of $30-10^3\,\msun$. The second
suggestion that the Galactic central complex IRS~13E consists of an IMBH with few 
stars still bound to it has also been ruled out \citep{Muzic08}. 

Additional circumstantial evidence comes from the observations that clusters with high
central concentration index ($c=\log{r_{tidal}/r_{core}}$) appear to have retained 
their low-mass stars, when compared to low-concentration clusters that show serious depletion.
A cluster loses low-mass stars due to evaporation, and mass segregation plays an important
role in this process by pushing the low-mass stars to the periphery of the cluster. However, it
is expected that IMBH in the globular cluster centre would quench mass segregation 
\citep{Gill08}. The degree of mass segregation present was already applied to rule out 
the presence of an IMBH in a low-concentration, small Galactic globular cluster 
NGC~2298 \citep{Pasquato09}, which is also heavily depleted in low-mass stars 
\citep{DeMarchi07}. Incidentally, NGC~2298 has a large core, which reinforces the 
conclusions of \citet{Hurley07} that a large core is not 
necessarily an indication of the presence of an IMBH in a globular cluster. 
 
Thus, if the formation of a central black hole is a normal stage in the life of a 
globular cluster, current observations favour that its presence would aid the survival of the 
globular cluster in the field of the Galaxy. 
 
\subsection{Model predictions; deviations from the log-linear relation}

Though the GRG sample statistically favours a log-linear relation, \cite{Barth05} 
noticed that it is possible that the low-mass AGN sample does flatten the $\msigma$ 
curve somewhat. They warn that while it is still too early for any 
definitive conclusion,  it may be that AGNs follow a different $\msigma$ relation altogether.
However, the sample of \citet{GreeneHo06} completely rules out the steepening of the slope 
\citep{Barth05}. We have applied log-square and log-cubic fits to our 
full sample in addition to the log-linear fit, and have calculated the AIC and BIC 
factors for these fits (using \citet{RProject}).\footnote{\footnotesize 
The idea behind the  Akaike Information Criterion \citep[AIC,][]{Akaike74} and 
Bayesian Information Criterion \citep[BIC,][]{Schwarz78} is that we expect the 
model with more parameters (e.g, log-quadratic) to achieve a higher maximized 
log-likelihood that the model with fewer parameters (log-linear model). However, 
it may be that the additional increase in the log-likelihood statistic, achieved 
with more parameters, is not worth adding the additional parameters. 
The smaller AIC values for the log-linear fit indicate that linear fit is better,
that is worth the additional parameters. However, since AIC is known to tend to overfit 
sometimes, i.e., it may favour models with more parameters than they 
should have, we have also calculated the BIC criteria, which imposes a larger penalty 
for increase in the number of parameters than AIC.}
Table~\ref{tab:fits} shows the values for these parameters. It can be can seen from 
the Table that log-linear relation provides a better fit to the $\msigma$ relation 
for the full sample. It should also be noted that the possible (low-significance) 
log-quadratic fit to GRG sample, reported in \citet{Gultekin09} and shown by a 
curved line in the left panel of Fig.~\ref{fig:msigma}, 
is ruled out once low-mass AGN and globular clusters are taken into account.

\begin{table}[h!]
\begin{center}
\caption{}
\label{tab:fits}
\vskip 0.1in
\begin{tabular}{|l|c|c|}
\hline
Full sample & AIC& BIC  \\
\hline
Log-linear$^{*}$       & 83.857 & 90.37 \\
Log-quadratic$^{\dag}$ & 85.518 & 94.22 \\ 
Log-cubic$^{\ddag}$    & 85.747 & 96.62 \\
\hline
\multicolumn{3}{l}{ } \\
\multicolumn{3}{l}{$^{*}$ $y=\a +\b x$, see Eq.~\ref{eq:basic},}\\
\multicolumn{3}{l}{$^{\dag}$ $y=\a +\b_1 x +\b_2 x^2$,}\\
\multicolumn{3}{l}{$^{\ddag}$ $y=\a +\b_1 x +\b_2 x^2 +\b_3 x^3$.} \\
\end{tabular}
\vskip 0.06in
\end{center} 
\end{table}

The remarkable consistency of the $\msigma$ correlation can be used as 
a constraint on cosmological models making specific predictions for the low-mass 
end of the $\msigma$ relation. The low-mass cut-off in the BH mass function today
provides a constraint on the mass function of seed black holes for the SMBH. 
If BH seeds are formed from the remnants of Population III stars and have masses
of $\sim 100\,\msun$, the so-called `light-seed model', 
the merger tree calculations predict high BH occupation
fractions in low-mass galaxies \citep{Volonteri08}, 
though with such a steep slope of $\msigma$ relation that
some galaxies hosting very tiny ($<10^4\,\msun$) black holes. The steepening of 
the $\msigma$ slope after $\sigma \sim 150$~km/sec is also predicted in the 
hierarchical black hole growth scenario by \citet{Granato04}, where 
supernova feedback becomes more efficient at slowing the AGN fueling rate. 

Another prediction, proposed by \citet{Wyithe06},
was that this relation should flatten in the range between
70 to 380 km/sec, in other words, a log-quadratic relation providing a better fit 
to the sample of SMBH masses and velocity dispersions. One of their predictions 
was that in bulges with $\sigma \sim 10$ km/sec the minimum mass of the 
black hole shall be $\sim 10^5 \msun$. The same flattening (at $20-50$ km/sec)
is predicted in the 'heavy-seed' model \citep{Volonteri08}. The different 
`seed' cases show markedly different behaviour for the predicted BH occupation
fraction and $\msigma$ relation at low masses, thus making globular clusters an 
ideal laboratory to differentiate between the models. 

\section{Conclusions}

We extend the $\msigma$ relationship for galaxies to lower black hole masses 
using black holes discovered in globular clusters.
The reported masses of black holes in the centres of globular clusters 
M15, 47~Tuc, $\omega$~Cen, NGC~6388 and G1 are consistent with the linear extrapolation 
of the $\msigma$ relationship for galaxies to the low-mass end.
Using this extrapolated relationship, we have estimated the masses of putative 
black holes in a sample of globular clusters, and find that these clusters 
are consistent with the $\mbh-$luminosity relationship as well. 

In the extended $\msigma$ plot, points corresponding to different types of 
stellar systems occupy distinct regions, suggesting that black hole masses of 
\abt~few~$\times~10^{4}~\msun$ represent the {\it lowest} limit for the central 
black holes  of galaxies.  Masses of the central black holes that are below this limit 
correspond to the globular cluster domain (keeping in mind that G1 and 
$\omega$~Cen are believed to be tidally stripped dwarf galaxies and not 
genuine globular clusters). 

The consistency of black hole masses in globular clusters with the 
extrapolated $\msigma$ and $\mbh-$luminosity relationships reinforces 
the idea that globular clusters harbour IMBHs in their centres. 
The central black holes of globular clusters place even stronger constraints 
than low-luminosity AGNs on theoretical models of supermassive black hole growth 
and evolution. If black holes in globular clusters do exist, they will rule out 
models that predict either steepening or flattening of the $\msigma$ relation
the low-mass region.

\appendix
\section{Appendix. Additional galaxies to the GRG sample.}
\label{AppA}

\begin{table*}[hb!]
\begin{center}
\small
\caption{Additional galaxies}
\label{tab:addgalaxies}
\vskip 0.1in
\begin{tabular}{|c|c|c|c|c|c|}
\hline
\rule[-3mm]{0mm}{8mm}
Galaxy &Type    & $\mbh$(high,low) & $\s_{\rm e}$ & $M_{\rm V}$&Method\\
          &         &  ($10^6\,\msun$)& (km/sec) &    &  \\ 
\hline
\rule[-3mm]{0mm}{8mm}NGC2974 &  E4,Sy2 & $170.0(190.,150.)$ & $236\pm 12$&-21.09&s \\   
\rule[-3mm]{0mm}{8mm}NGC3414 &S0 & $250.0(280.,220.)$ & $205\pm 10$ &-20.25& s    \\
\rule[-3mm]{0mm}{8mm}NGC4395& Sm, dwarf& $0.36(0.47,0.25)$&$30\pm 5$  & -11.00&r \\
\rule[-3mm]{0mm}{8mm}NGC4552  & E,L & $500.0(550.,450.)$& $263\pm 13$&-21.65&s\\
\rule[-3mm]{0mm}{8mm}NGC4621 &E5  & $400.0(440.,360.)$& $231\pm 12$&-21.74&s\\
\rule[-3mm]{0mm}{8mm}NGC5813 & E1,L & $700.0(770.,730.)$& $237\pm 12$&-22.01&s\\
\rule[-3mm]{0mm}{8mm}NGC5846 & E0 & $1100.0(1200.,1000.)$& $238\pm 12$&   &s\\
\rule[-3mm]{0mm}{8mm}NGC7332 & S0 & $13.0(19.0,8.0)$& $125\pm 16$&-19.62&s\\
\rule[-3mm]{0mm}{8mm}IC2560& SBb,Sy2 & $2.9(3.5,2.3)$& $137\pm 14$&-20.7&m\\
\rule[-3mm]{0mm}{8mm}POX52 & dE,N & $0.14(0.25,0.03)$& $36\pm 5$&-17.6&multiple\\
\hline\end{tabular}
\end{center}
\noindent
{\it Key to columns:} (1) name of the galaxy, (2) Hubble type of the galaxy and activity of the 
nucleus if present (Sy1 = type 1 Seyfert; Sy2 = type 2 Seyfert, L = LINER), (3) mass of the 
central black hole (high, low), (4) effective stellar velocity dispersion of the bulge,
(5) absolute V-band luminosities are taken from \cite{Lauer07} with the following 
exceptions: NGC~4395 \citep{Peterson05}, IC2560 \citep{Hunt04} and POX52 \citep{Barth04}; 
(6) detection method (s = stars, g = gas, m = H$_2$0 masers, r = reverberation mapping).\\ 
\end{table*}

\small
\appendix
\section*{Appendix B. Candidate Globular Clusters.}
\label{AppB}
\begin{table*}[h!]
\begin{center}
\caption{Globular clusters with predicted masses for central black holes.}
\label{tab:GCmasses}
\begin{tabular}{|l|c|l|l|c|l|l|}\hline
\rule[-3mm]{0mm}{8mm}Cluster &Other& $\mbh$($10^3\,\msun$)& $\s$(km/sec) 
& $M_{\rm V}$ & Reason & 0ther predictions ($\msun$)\\
\hline
NGC362  &     & $73.9\pm  1 0.2$   &$6.2\pm 3.0$ [1] &-8.41&c? & \\
NGC1851 &     & $917.0\pm   1.7$   &$11.3\pm 2.5$ [1] & -8.33&HVS & \\  
NGC1904 & M79 & $35.3\pm    3.1$   &$5.2\pm 1.04$ [3]&-7.86&c? & \\      
NGC2808 &     & $1874.7\pm  2.8$   &$13.4\pm 2.68$ [2] &-9.39&HVS,radio,EHB &550[2],$110$-$3100$[3] \\
NGC5286 &     & $291.6\pm  10.9$   &$8.6\pm 4.3$ [1] &-8.61&B. & \\
NGC5694 &     & $69.0\pm   3.14$   &$6.1\pm 1.3$ [1]&-7.81&B. & \\ 
NGC5824$^{\dag}$& &$850.8\pm  2.3$ &$11.1\pm 1.6$ [1]& -8.84& c?,HVS,B. & \\
NGC5904 & M5  & $90.1\pm   10.6$   &$6.5\pm 3.2$ [1]&-8.81& radio & 320[2]\\
NGC6093 & M80 & $2610.4\pm  9.7$   &$14.5\pm 7.0$ [1] & -8.23&HVS,EHB,B. & $1600$[1],$1000$-$1200$[3]\\
NGC6205 & M13 & $130.5\pm   2.9$   &$7.1\pm 1.42$ [3]&-8.70&radio,EHB& 250[2],$130$-$2400$[3]\\
NGC6256 &Ter12& $96.1\pm   12.0$   &$6.6\pm 3.4$ [1]&-6.52&c& \\
NGC6266 & M62 & $3360.7\pm  9.6$   &$15.4\pm 7.4$ [1] & -9.19&c?,HVS,radio,EHB,B.& $3000$[1],450[2],290-1500[3]\\
NGC6273 &     & $1335.8\pm  2.0$   &$12.36\pm 1.24$ [6]&-9.18&radio& 410[2]\\
NGC6284 &     & $108.9\pm  11.1$   &$6.8\pm 3.4$ [1]&-7.97&c& \\
NGC6293 &     & $238.8\pm  11.7$   &$8.2\pm 4.2$ [1]&-7.77&c& \\
NGC6325 &     & $84.4\pm   12.1$   &$6.4\pm 3.3$ [1]&-6.95&c& \\
NGC6333 & M9  & $172.7\pm   2.2$   &$7.59 \pm 0.76$ [6]&-7.94&c& \\
NGC6342 &     & $35.3\pm  11.22$   &$5.2\pm 2.6$ [1]&-6.44&c& \\
NGC6355 &     & $356.2\pm   2.1$   &$9.02\pm 0.902$ [6]&-8.08&c& \\
NGC6380 & Ton1& $77.5\pm   2.25$   &$6.27\pm 0.63$ [6]&-7.46&c? & \\
NGC6397$^{\dag}$& &$19.3\pm 2.38$  &$4.5\pm 0.45$ [6]&-6.63&c,B.,radio&390-1290[10] \\
NGC6402 &      & $238.8\pm  2.9$  &$8.2\pm 1.64$ [3]&-9.12&radio&390[2] \\
NGC6440 &      & $352.9\pm  3.1$  &$9.0\pm 2.0$ [7]&-8.75&radio& 300[2]\\
NGC6441 &     & $549.1\pm  2.8$ & $10.0\pm 2.0$ [7] &-9.64&HVS,radio& 470[2]\\
NGC6453 &     & $114.3\pm  2.2$    &$6.88 \pm 0.68$ [6]&-6.88&c& \\
Ter 6       &HP5& $131.3\pm  2.2$   & $7.11\pm 0.7$[6]&-7.67&c & \\
NGC6522 &     & $146.6\pm  9.9$    &$7.3\pm 3.5$ [1]&-7.67&c& \\
NGC6541$^{\dag}$& &$238.8\pm  3.6$&$8.2\pm 2.1$ [2]&-8.37&c?,B.& \\  
NGC6544 &     & $364.6\pm  2.1$    &$9.07\pm 0.91$ [6]&-6.66&c?& \\
NGC6558 &     & $6.7\pm  12.4$    &$3.5\pm 1.8$ [1]&-6.46&c& \\
NGC6624 &     & $41.4\pm  2.3$    &$5.4\pm 0.54$ [1]&-7.49&c& \\
NGC6626 &  M28& $242.5\pm  2.5$    &$8.23\pm 1.28$ [2]&-8.18&radio& 210[2]\\
NGC6642 &     &  $10.1\pm  2.4$   &$3.86\pm 0.39$ [6]&-6.77&c?& \\
NGC6656 & M22 & $5075.5\pm  1.9$    &$16.99\pm 1.7$ [5]&-8.50&radio& $240$[2] \\
NGC6681 & M70 & $549.1\pm  3.56$    &$10.0\pm 2.6$ [1]&-7.11&c& \\
NGC6715 & M54 &$10491.2\pm 1.64$  & $20.2\pm 0.7$ [8]&-10.01&HVS,radio,EHB,B. &960[2],$700$-$3200$[3],9400[9]\\
NGC6752 &     &$1354.0\pm  1.80$ & $12.4\pm 0.5$ [4] &-7.73& c,HVS,PSR &500-1000[4] \\
NGC7099 & M30 & $48.2\pm  2.3$    &$5.6\pm 0.56$ [2]&-7.43&c & \\
\hline
\end{tabular}
\end{center}
\small
{\it Key to columns:} (1)-(2) cluster identification number and other name,
(3) calculated BH mass with error bars, (4) central velocity dispersion with reference; 
when the error on dispersion was not given, we assumed it at $10\%$, 
(5) V-band absolute magnitude, (6) references to the reasons of selection, (7) 
masses predicted in the literature.\\
$^{\dag}$ Not likely to host a black hole according to \cite{Baumgardt05}; 
belongs, however, to the ``c'' set.\\
{\it References to velocity dispersions}: (1) \cite{Dubath97}, 
(2) \citet{PryorMeylan93}, (3) \citet{McLaughlinvanderMarel05},
(4) \citet{Drukier03} (5) \cite{Chen04} (6) \cite{Webbink85}, 
(7) \citet{Origlia08}; (8) \citet{Ibata09}\\
{\it References to the reasons of selection}: ``c''=post-core-collapse (pcc) 
morphology; ``c?''=possible p.c.c. \citep{Trager95,Lugger95}; 
``HVS''=high velocity stars \citep{DrukierBailyn03}; 
``radio''=radio observations \citep{Maccarone04,Bash08}; ``EHB''=extreme horizontal 
branch stars \citep{Miocchi07}; ``PSR''=detection by pulsar dynamics 
\citep{Colpi03}; ``B.''=Baumgardt's sample \citep{Baumgardt05}. \\
{\it References to other predictions}: [1] \citet{Bash08}: mass estimates from $\msigma$
relation; 
[2] \citet{Maccarone04}: mass estimates from $\mbh=0.1 M_{\rm cluster}$;
[3] Miocchi (2007): mass estimates from the observed
standard concentration parameter and slope of the central surface-brightness profile;
[4] \cite{Colpi03}: mass range inferred from dynamical estimates;
[9] Ibata et al. (2009): this mass is the best fit from observations of velocity and
density cusps; [10] \citet{deRijcke06}: mass estimate from $3\sigma$ upper limit on radio
emission.
\end{table*}

\bibliographystyle{apj}
\bibliography{SafonovaShastri}

\end{document}